    \documentclass[fleqn,twoside]{article}
    \usepackage{espcrc2}
    \usepackage{graphicx}
    \usepackage{amssymb}
\newcommand{\AmS}{{\protect\the\textfont2
  A\kern-.1667em\lower.5ex\hbox{M}\kern-.125emS}}

\title{Exchange of high and low energy photons in 
ultraperipheral relativistic heavy-ion collisions}

\author{G.Baur\address{Institut f\"ur Kernphysik, 
        Forschungszentrum J\"ulich\\ 
        D-52428 J\"ulich, Germany}%
}
       
\begin{document}

\begin{abstract}

Ultraperipheral collisions at collider 
energies are a useful tool to study photon-hadron
(proton/nucleus) and photon-photon
interactions in a hitherto unexplored energy 
regime. Theoretical tools to study these processes
are briefly described. Some current results and problems are discussed.

\end{abstract}

% typeset front matter (including abstract)
\maketitle

\section{Introduction}
Hadron-hadron collisions can be an interesting tool to study 
photon-photon and photon-hadron interactions.
Of course, hadron-hadron collisions are
dominated by strong interactions between the hadrons.
However, by choosing collisions with large 
impact parameter b (or, equivalently, small momentum transfer)
one can suppress these strong interactions
(so-called ultraperipheral collions (UPC)).
An early observation of  dimuon pairs at the ISR without 
the production of hadrons is described by F.Vannucci \cite{van}.
Dielectrons from ultraperipheral peripheral interactions at the 
Tevatron were discussed at this workshop \cite{pin}.

A fast moving charged particle acts like 
a spectrum of quasireal photons.
This equivalent photon spectrum is soft, with
a maximum photon energy increasing with beam energy.
In heavy ion collisions this field is so strong that 
there is multiple photon exchange and cross sections,
especially for soft processes, can be huge. 

In Section 2 the characteristic properties of the equivalent photon
spectrum are recalled. Glauber or semiclassical models are
used to treat multiphoton processes.
Vector meson production at RHIC energies is discussed in 
Section 3. An especially interesting feature is the 
transverse momentum distribution. 
Section 4 is devoted to the special problem of 
$e^+ e^-$ pair production. A very brief outlook on 
prospects at the LHC is given in Section 5, 
conclusions are given in Section 6.
There are various  reviews 
which reflect the progress of the field during the last decades
\cite{ber,soff,bau,cms,bns,ect,photon2007,yr}.

\section{Theoretical description of UPC}
\subsection{One photon exchange, equivalent photon aproximation}
\begin{figure}[h]
\centerline{\includegraphics[width=1.0\columnwidth]{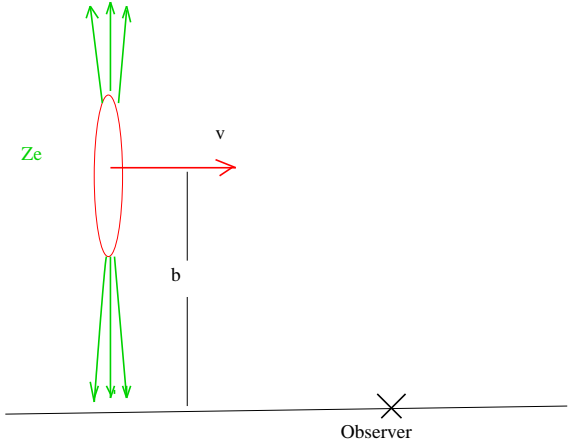}}
\caption{A fast charged particle 
moving on a straight line with impact parameter b causes a time-dependent
electromagnetic field at the point of the observer.
This field corresponds to a spectrum of equivalent photons.}\label{Fig:PH}
\end{figure}

The time-dependent electromagnetic field of a fast moving charged particle
can be thought of as a spectrum of (quasireal, or equivalent)
photons~\cite{fermi}, see Figure \ref{Fig:PH}.
The determination of the 
equivalent (or Weizs\"acker-Williams) photon spectrum
corresponding to a fast particle moving past an observer on a 
straight line path with impact parameter $b$ is a textbook example
~\cite{jac}. 

The probability $P(b)$ of a specific
photon-hadron reaction to occur in a collision
with an impact parameter $b$
is given by $P(b)=N(\omega, b) \sigma_{\gamma h}(\omega)$,
where $\sigma_{\gamma h}$ is the corresponding photoproduction
cross section. The equivalent photon spectrum can be calculated
analytically, a useful approximation for qualitative considerations is 
\begin{equation}
N(\omega, b)=\frac{Z^2 \alpha}{\pi^2 b^2}
\label{Eq:nb}
\end{equation}
for $\omega<\frac{\gamma}{b}$
and zero otherwise. The nuclear charge is given 
by $Z$, heavy ions have particularly high photon fluxes,
however, this is partially offset by the lower ion-ion luminosities,
as compared to the p-p case.

%\begin{wrapfigure}{r}{1.0\columnwidth}
%\end{wrapfigure}
The impact parameter b is restricted to
\begin{equation}
b>b_{min} \sim R_1 + R_2
\end{equation}
where $R_1$ and $R_2$ denote the sizes of the hadrons.

For heavy ion scattering the Coulomb parameter
$\eta \equiv \frac{Z^2 e^2}{\hbar v}\sim Z^2/137$ is 
much larger than unity and it is in principle possible to determine the
impact parameter
by measuring the angle of Coulomb scattering.
Whereas this is experimentally feasible at lower 
($\sim GeV/A$) energies
~\cite{aum}, this angle is too small at collider
energies. So one generally measures quantities
integrated over all impact parameters.
Too small impact parameters are recognized since the event is
dominated by the violent strong interactions.

The photon spectrum Eq.~\ref{Eq:nb} extends up to 
a maximum photon energy given by 
\begin{equation}
\omega_{max}=\frac{\gamma}{b_{min}} .
\end{equation}
This energy is about 3 GeV at RHIC (Au-Au, $\gamma \sim 100$),
and 100 GeV at LHC (Pb-Pb, $\gamma \sim 3000$) in the collider
system. For p-p collisions at the LHC it is several hundreds
of GeV. With the possibitity to tag on these photons by observing the 
forward protons this is of great interest, as was discussed at this
workshop.
\subsection{Improvement on equivalent photon approximation}
As the name implies, the equivalent photon aproximation
is an approximation. The virtual, spacelike photon which is 
exchanged is assumed to be real, and a suitable cut-off has to be introduced.
This cut-off parameter enters only in a logarithm, and it is not sensitive
to details.
It is of course also possible to calculate the one-photon exchange process 
in a direct way by evaluating
the corresponding Feynman graph, where the exchanged 
photon is treated as a virtual particle. 
It is straightforward to introduce electromagnetic formfactors
for the protons or nuclei. They are suffiently well known
for the present purposes. This could 
be directly implemented into computer programs like Madgraph (see Michel
Herquert, this workshop). The approximations introduced in the 
equivalent photon approximation are discussed in detail in 
\cite{budnev}, see also \cite{hama}.

\subsection{Multiphoton processes, Glauber approximation}
%% section headers !

\begin{figure}[h]
\centerline{\includegraphics[width=1.0\columnwidth]
{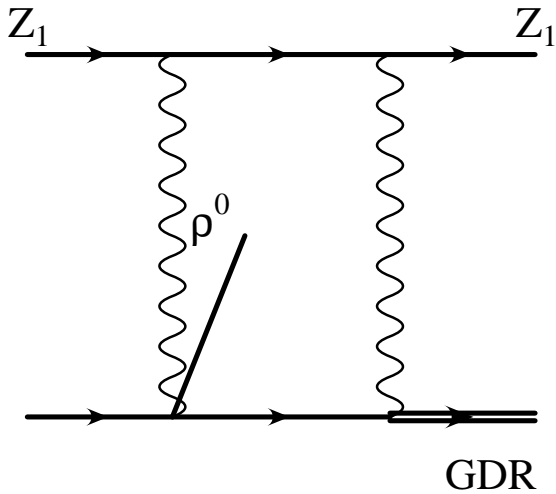}}
\caption{A graph contributing to the simultaneous 
production of a $\rho$-meson and the excitation of the 
giant dipole resonance (GDR).}\label{Fig:mua}
\end{figure}

For heavy ions the probability of an electromagnetic
interaction in ultraperipheral collisions is especially 
large, and multiphoton processes occur, see e.g.~\cite{npa}.
We mention $e^+e^-$ pair production where 
the impact parameter dependent total pair
production probability $P(b)$
is of order unity. Multiple pairs can be produced,
however they may be hard to detect due to their low
transverse momentum.
The nuclear giant dipole resonance is excited with  
probabilities of the order of one third.
For a reliable estimate of the giant dipole excitation 
probability on can use the global formula
\begin{equation}
P(b)=5.4 \times 10^{-5}Z^3NA^{-2/3} fm^2
\end{equation}
where a collision of two identical nuclei with mass number A
and neutron number N was assumed.
It is based on the Thomas-Reiche-Kuhn sum rule, and
the excitation energy of the giant dipole resonance 
is assumed to be $80 A^{-1/3}$MeV. For more accurate numbers
one uses the experimentally known photonuclear cross sections
and folds them with the equivalent photon spectrum. 
In Figure \ref{Fig:mua} one of the
graphs is shown which leads to the electromagnetic 
production of a $\rho^0$ along with the excitation 
of the giant dipole resonance. 
These graphs can conveniently be evaluated in semiclassical or eikonal
theories~\cite{npa}.
Whereas the total cross section for one-photon exchange 
processes depends only logarithmically on the minimum
cut-off radius R, the N-photon exchange process scales as
$1/R^{2(N-1)}$. Thus it becomes more important to 
have a reliable value for this strong cut-off parameter
in the case of multiphoton exchange.

%\begin{figure}[h]
%\centerline{\includegraphics[width=1.0\columnwidth]{fig3b.ps}}
%\caption{A fast (relativistic charged particle...}\label{Fig:mub}
%\end{figure}
%Captions of figures and tables appear {\em below} the figure/table.
%When referring to Figure~\ref{Fig:PH} capitalize the first letter.

The giant dipole resonance decays dominantly into a neutron
and a residual nucleus.
The neutron can be  detected in the forward direction. This 
can serve as a luminosity monitor  and also as 
a trigger on UPC \cite{mchiu}.

\section{Vector meson production at RHIC}

Recently $\rho^0$ photoproduction in ultraperipheral relativistic heavy ion 
collisions at $\sqrt{s_{NN}}$=200 GeV was reported by the 
STAR collaboration \cite{abe}.
Two kinds of triggers were used, a topology trigger and a 
trigger based on the observation of neutrons in the Zero
Degree Calorimeter (ZDC). In the latter case, the vector meson 
photoproduction is accompanied by mutual electromagnetic
excitation of the giant dipole resonance. I would like to point out that
a reliable theoretical value for the giant resonance excitation 
probability (or rather photoneutron cross section) 
has to be used in order to extract absolute 
values for the differential photoproduction cross section.

A feature unique to photoproduction 
in hadron-hadron collisions is an interference effect
~\cite{kn}: a vector meson can be produced by a photon
originating from either of the hadrons. 
It was shown in~\cite{kn} that this interference effect 
leads to a reduction of the transverse momentum spectrum
of the vector mesons for small transverse momenta.
Another theoretical approach~\cite{hbt} 
leads to very similar conclusions. 
In this paper Glauber and semiclassical methods
(shown to be essentially equivalent) are used to describe 
the interference phenomenon. The differential cross section is found to be
\begin{equation}
\frac{d^3\sigma}{d^2v_\perp dY}=\frac{1}{2(2\pi)^3} \sum_{e_V} 
\int |a|^2 d^2b
\end{equation}
where
\begin{eqnarray}
a=a_1(b) a_2(b) (a_V(\vec b,\vec v_\perp, Y)+
e^{-i\vec v_\perp \cdot \vec b}a_V(-\vec b, \vec v_\perp,-Y))
\end{eqnarray}
Here $a_1$ and $a_2$ denote the nuclear excitation,
the amplitude for vector meson photoproduction 
is denoted by $a_V$. The vector meson perpendicular momentum is $\vec v_\perp$,
Y is the rapitidy and $\vec b$ the impact parameter.
(Preliminary) 
experimental results from STAR/RHIC indeed show a dip
for small transverse momenta, see e.g. Ref. \cite{yr}.

\section{Electron-positron pair production}
Electron-positron pair production in ultrarelativistic
heavy ion collisions was recently reviewed in \cite{bht07}.
The electromagnetic fields are very strong, of the order of  
or even larger than the Schwinger critical field strength which is given by 
\begin{equation}
E_c=\frac{m_e^2c^3}{e \hbar} \simeq 1.3 \times 10^{16} \rm V/cm
\end{equation}
where $m_e$ denotes the electron mass.
 However, these
fields act only for a very short time and perturbation theory 
is an appropriate tool to study these processes. 
Higher order effects are present and were studied theoretically 
over the past years, and the effects are well understood.
For further details and a comparison to the physics of ultraintense laser
pulses we refer to \cite{bht07}. 

Electron-positron pair production in ultraperipheral collisions 
accompanied by mutual giant dipole excitation was measured at STAR  and analysed with 
equivalent photon method \cite{adams}. A lowest order QED
(two-photon pair production) calculation is given in \cite{hbt04}. 
In view of the limited statistics, agreement between theory and experiment 
can be considered as satisfactory.

In a more recent analysis by A. Baltz \cite{bal08}
the  neutron production probability is treated in more detail
and higher order QED effects are included. 
The author concludes that 
the experiment \cite{adams} is in good agreement
with the higher order calculation. 

\subsection{Bound-free pair production}
In bound-free pair production the electron is produced
in a bound atomic orbit (K-,L-,..shell).
  It scales approximately as 
\begin{equation}
\sigma \sim \frac{Z^7 ln \gamma \delta_{l0}}{n^3}
\end{equation}
where n and l denote the principal and agular momentum
quantum numbers of the atomic bound state. 
These ions with their changed charge-to mass ratio will get
lost from the beam and they will heat up the beam pipe
in a hot spot.
It was
identified as a serious limit for the luminosity in Pb-Pb collisions at LHC.
The bound-free pair production cross section was recently measured at RHIC \cite{bruce}.
Agreement with theory \cite{helmar} is good.
The bound-free pair production mechanism was also used in the production of 
fast antihydrogen at LEAR in 1996 by W. Oelert et al.\cite{antihy}.

\section{Opportunities for UPC at LHC}

The events caused by  ultraperipheral collisions 
may be called 'silent events', as compared to the 
voilent central collisions. However, the collision of 
a 100 GeV equivalent photon with another one of similar 
energy can lead to lead to events that are not so 'silent'.
Anyway, the events are there, but experimentalists must find ways to trigger on them.

The maximum photon energy scales linearly with 
the Lorentz factor $\gamma$, see eq. 3. This leads to
a significant widening of the opportunities at LHC as compared to
RHIC. A most promising area is low-x QCD studies.
The experiments at HERA have shown that 
photoproduction processes provide a well-understood
probe of the gluon density in the proton. At LHC,
such processes could be extended to invariant
$\gamma p$ energies exceeding the maximal HERA energy
by a factor of 10. This would allow to use dijet (charm, etc.)
production to measure the gluon density in the proton
and/or nucleus down to $x \sim 3 \times 10^{-5}$.
 Ultraperipheral collisions would also allow one 
to study the coherent production of heavy quarkonia,
$\gamma + A \rightarrow J/\Psi (\Upsilon) + A$
at $x \lessapprox 10^{-2}$, and to investigate the 
propagation of small dipoles through the nuclear medium
at high energies, see Ref. \cite{fra}, see also  
Refs.~\cite{kopel,macha}. 
Dijet production via photon-gluon fusion is 
calculated in Ref. \cite{svw}. Very large rates 
are obtained that will considerably extend the HERA x range.

In addition to diffractive processes in proton-proton 
collisions at LHC also a rich program of proton-photon 
and photon-photon physics can be pursued, see Ref. \cite{cern}.
The photon flux is lower as compared to the 
heavy ion case due to the $Z^2$-factor, but this
is at least partly compensated by higher beam luminosities.
The photon spectrum is harder due to the smaller size as compared to 
the heavy ions, this leads  to a lower value of $b_{min}$ in Eq. 3 . 
Possibilities for electroweak physics and
beyond were discussed at the present workshop.
Tagging on photon energy by measuring the energy loss of 
the scattered protons in the forward detector TOTEM
is an important feature.

The production of the Higgs boson by photon-photon fusion would
be of obvious interest. A value of 3.9 nb for Au-Au collisions at the 
LHC was given recently in \cite{lemi}. This is a large value, as
compared to numbers reported previously, see e.g. Ch. 7.2.7 of
\cite{cms}. Absorption seems to be treated somewhat differently 
in \cite{lemi} as compared to the methods developed in \cite{bafe,caja}.

\section{Conclusions}
Ultraperipheral collisions at the LHC can provide a new means of
studying small x QCD, electroweak physics and 
physics beyond the standard model. Some issues 
related to the theoretical description of ultraperipheral 
heavy ion collisions are briefly discussed
in this contribution.

\section*{Acknowledgments}

I would like to thank David d'Enterria, Michael Klasen and 
Krzystof Piotrzkowski for their kind 
invitation to this timely and stimulating workshop.
I am grateful to  Carlos Bertulani, Kai Hencken, and Dirk Trautmann
for their collaboration over many years. I also wish to thank 
Spencer Klein and Joakim Nystrand for discussions.

\end{document}